# The dichotomy structure of Y chromosome Haplogroup N

Kang Hu[1]☯, Shi Yan[2,]*☯, Kai Liu[4], Chao Ning[5], Lan-Hai Wei[2], Shi-Lin Li[2], Bing Song[2], Ge Yu[2], Feng Chen[1], Li-Jun Liu[1], Zhi-Peng Zhao[1], Chuan-Chao Wang[2], Ya-Jun Yang[2], Zhen-Dong Qin[2], Jing-Ze Tan[2], Fu-Zhong Xue[2], Hui Li[2], Long-Li Kang[1]*, Li Jin[2,3]

[1] Key Laboratory of High Altitude Environment and Gene Related to Disease of Tibet Ministry of Education, Tibet University for Nationalities, Xianyang, Shaanxi, China
[2] Ministry of Education Key Laboratory of Contemporary Anthropology and Center for Evolutionary Biology, School of Life Sciences and Institutes of Biomedical Sciences, Fudan University, Shanghai 200433, China
[3] Chinese Academy of Sciences Key Laboratory of Computational Biology, CAS-MPG Partner Institute for Computational Biology, SIBS, CAS, Shanghai 200031, China
[4] Tibet Occupational College of Technology, Lhasa, Tibet 850000, China
[5] College of Life Science, Jilin University, Changchun, Jilin 130012, China

☯These authors contribute equally to this article

*Correspondence to: S.Y. (biopolyhedron@gmail.com) or L.L.K. (klonglister@gmail.com)

## Running Title:

Dichotomy of N

## Abstract:

Haplogroup N-M231 of human Y chromosome is a common clade from Eastern Asia to Northern Europe, being one of the most frequent haplogroups in Altaic and Uralic-speaking populations. Using newly discovered bi-allelic markers from high-throughput DNA sequencing, we largely improved the phylogeny of Haplogroup N, in which 16 subclades could be identified by 33 SNPs. More than 400 males belonging to Haplogroup N in 34 populations in China were successfully genotyped, and populations in Northern Asia and Eastern Europe were also compared together. We found that all the N samples were typed as inside either clade N1-F1206 (including former N1a-M128, N1b-P43 and N1c-M46 clades), most of which were found in Altaic, Uralic, Russian and Chinese-speaking populations, or N2-F2930, common in Tibeto-Burman and Chinese-speaking populations. Our detailed results suggest that Haplogroup N developed in the region of China since the final stage of late Paleolithic Era.

## Introduction

High-throughput sequencing has brought great amount of new information to the research of history and prehistory of human populations. Since the Y chromosome is strictly passed on from father to son in a population, the phylogeny of Y chromosome commonly shows parallelism to the population history, therefore the phylogeny can provide us hints for tracing the past of human populations that was not written in the historical literatures [1].

In Eastern and Northern Eurasia, the Haplogroup N-M231 of Y chromosome might be a key to understand the early development of the Altaic, Uralic, and Sino-Tibetan populations. From the studies of ancient DNA, Haplogroup N appeared as one of the most frequent clades in the late Neolithic cultures along the Great Wall in China [2], and therefore might be a major component of the eminent nomad peoples in the Steppe, like the Xiongnu, Xianbei, Tujue (Turkic Khaganate), and Huihu (Uyghur Khaganate). Nowadays, this haplogroup is widespread in a great range of the Eurasian continent, from Southeast Asia unto Northern Europe [3]. The highest frequency was found in several Uralic, Indo-European, and Altaic speaking populations in Siberia and eastern Europe, e.g. Finn (42 – 71%), Baltics (42 – 44 %), Russian (6 – 54%), and Yakut (79 – 100%), as well in several Sino-Tibetan speaking populations on the southeastern fringe of the Tibetan Plateau, like Yi (11 – 32%) and Lhoba (35%) [4-14]. Haplogroup N is also dispersed in lower frequency in many other Altaic-speaking ethnic groups (0 – 20%), as well as Han Chinese (2 – 9%) [15] and Tibetan (2 – 15%) [16].

Many studies have worked with the subclades N-M178 and N-P43 (N1c1 and N1b, according to the YCC 2008 tree [17]), which are prevalent in Europe. From the those studies, Rootsi et al. suggested a counter-clockwise route of Haplogroup N with the origin from China, and serial bottlenecks in Siberia, and secondary expansions in eastern Europe [3]. However, till now, little was known about the deep clades of Haplogroup N, due to the fact that few SNPs were available for genotyping, so many samples of Haplogroup N could only be classified as N*-M231(xM128,P43,M46) or N1*-LLY22g(xM128,P43,M46), most of which are located inside China. Therefore, investigation of the upstream phylogeny of Haplogroup N would be very informative for the early history of China, as well as the Eurasian Steppe and Eastern Europe.

By hybridization capturing and Illumina sequencing of 9 Han Chinese male Y chromosomes of Haplogroup N (including several individuals belonging to the paragroup N*-M231(xM128,P43,M46)), more than 300 new SNPs (with "F" as prefix) were found, and a renewed phylogenetic tree was drawn [18], in which the most recent common ancestor of the sequenced samples are dated at 15.8 kya (95% CI: 13.8 – 18.0). To better understand the genealogy, the geographic distribution, as well as the origin of the entire Haplogroup N, in this study, we genotyped more than 400 samples of this haplogroup from 34 populations in China with selected 33 SNPs.

# Methods

## Sample collection

In this study, blood or buccal swab samples of 34 populations in China were collected, including Sino-Tibetan-speaking Han, Hui (Chinese-speaking ethnic group with Central and Western Asian mixture), Tibetan, Sherpa, Mönpa, Deng, Lhoba, Jiarong, Queyu, Muya, Qiang, and Tujia; Altaic-speaking Mongols; and Tai-Kadai-speaking Zhuang, with a total of more than 6,900 samples (Fig. 1). The sampling was under the approval of the Ethics Committee of Biological Research at Fudan University, with informed consent signed by the sample donors.

## Genotyping

DNA of the above mentioned male samples were first extracted, and were tested with SNP M231, then those samples with derived allele (n = 513) were further genotyped in this study. The 33 chosen SNPs were genotyped using SNaPshot multiplex kit (ABI, Carlsbad, California, US) [19] in 3 panels, and all the amplification primers and extension primers were newly designed in this study (SI Table 1). After multiplex PCR and single nucleotide primer extension, the SNP results were read from an ABI 3730 sequencer. Y-Filer kit (ABI, Carlsbad, California, US) was used for genotyping of 17 STRs (DYS19, DYS389a/b, DYS390, DYS391, DYS392, DYS393, DYS437, DYS438, DYS439, DYS448, DYS456, DYS458, DYS635, Y-GATA-H4, DYS385a/b).

## Partial deletion in AZFc

Partial deletion in AZFc was detected using multiple PCR of sY1201, sY1191 (showing b2/b3

deletion), sY1291 (showing gr/gr deletion), and sY1161. The primers' sequences and PCR condition followed the previous study [20].

## Results

Using the 33 bi-allelic markers, 16 distinct haplogroups that contain some individuals were discovered inside Haplogroup N (Fig. 1). Although 513 samples were categorized as members of Haplogroup N in previous genotyping, due to long preservation time, only 440 were successfully genotyped for downstream SNPs. The former N1-defining SNP LLY22g was removed from the phylogenetic tree due to recurrent mutation events (see Discussion). All the genotyped samples were categorized as either N1-F1206 or N2-F2930, while no N*(xF1206,F2930) samples were found. The three formerly known main subclades – N1a-M128, N1b-P43, and N1c-M46 were all found inside the new N1-F1206 clade, renamed as N1a1a1, N1a2, and N1b, respectively. M128 is now located downstream to the newly found SNPs F3163 (which is also upstream to P43), F1154, and F2759. The downstream structure of the formerly well-known clade M46+ (or Tat+) can now be further clarified. No M46+, M178- samples were found in this study, while F3331, F1419 or F4325, F3271, and F2288 (in the order from up- to downstream) each can separate some samples.

Since all the N samples could be divided into N1-F1206 and N2-F2930, the distribution of the two clades are drawn on the map (Fig. 2), and we found a clear trend, that N1 are more in the north and N2 in the south. All the genotyped N's in Mongolic-speaking populations belong to N1-F1206 clade, specifically, N1b1a-F1419 attributes to 60% of the whole population of Buriyad in Hulun Buir. On the other hand, the vast majority of N's in southern groups, mainly Tibeto-Burman (Bodic, Qiangic,

Jiarong, Tujia, Mishmi, and Tani) and Tai-Kadai (Zhuang) populations, belong to N2-F2930 clade. The Chinese-speaking Han and Hui showed an intermediate pattern, that N1 and N2 varies among regions, with a general gradient that N1 declines from north to south and N2 vice versa.

We compared the genotyping results of this study with numerous literatures (SI Table 2) [4-7,12,13,21-29], and checked the distribution of N's (Fig. 3). Since there were no genotyping of SNPs other than N1a1a1-M128, N1a2-P43, and N1b-M46 available in the literatures, we used the sum of the three subclades' frequency as the minimum of N1-F1206. We found that in accordance with the data in this study, the trend of north / south with N1 / N2 remains clear. Many populations north to China, including Tungusic, Uralic, Slavic, Baltic populations, as well as some Turkic speaking groups (Tuva, Yakut), showed that nearly all the N's belong to either N1b-M46 or N1a2-P43, and therefore under N1-F1206, while the southern ethnical groups, like Lolo-Burmese-speaking Yi and Lahu showed none of the three clades, and are possibly with high frequency of N2-F2930 (one Yi sample in this study was tested as N2a1-F1833).

From the STR data (SI Table 3), we can estimate the coalescence time of the subclades of Haplogroup N in the concerned populations. There have been long-standing debates lasting for more than 10 years about which STR mutation rate should be used for time estimation in history [30,31], despite of this, we pose here our time estimations for the reader's references, at least to show their relative age (Table 1). The age of each clade was calculated as the average STR-based time between the mean STR values (supposed as the ancestor node) and each modern member using Slatkin's method and father-son pair Y-STR mutation rate [32-34], and the dates were linearly calibrated to the divergence time of N1-F1206 and N2-F2930 at 15.8 kya (thousand years ago), which was achieved from the high-throughput sequencing of Y chromosomes [18]. We are aware that the estimated average coalescence time of a clade is for the sampled populations, rather than the divergence date of a whole clade, and the former time can be much younger than the latter. More convincing date estimation must be made after future high-throughput sequencing of more individuals of Haplogroup N.

Despite of those uncertainties, our calculation implies that N1 and N2 each has diverged already at the final stage of Paleolithic Era, shortly after the first divergence of N. On the other hand, several notable subclades expanded only a few thousand years before present, like N1a1-F1154 (most of which are N1a1a1-M128 and distribute in Han Chinese), which diverged ~5 kya, at the late Neolithic Era, as well as N1b1a1a-F3271 (here mainly Buriyad, and a few samples also in Bargud and Han Chinese), which diverged at only ~2 kya, probably reflecting a strong bottleneck in the history of Buriyad in China.

# Discussions

## Recurrent mutation of LLY22g

The SNP LLY22g (C/C > C/A) which defines former Haplogroup N1 was not genotyped for all the samples in this study. The both loci (chrY:24270184 and 24578321 of hg19) are located in the palindrome P3 of ampliconic region, with 2 highly similar copies [35,36]. As described before [18], although most of the N samples show C/A genotype on LLY22g, we found three genotypes, C/C,

C/A, and A/A inside the N2-F2930 clade. Therefore, we suggest that the LLY22g C/C > C/A occurred already before the split of N1 and N2, and the C/C genotype inside Haplogroup N was caused by reverse mutation, or rather likely a gene conversion between the both LLY22g loci, since recurrent de novo mutations are rarely found inside the human Y chromosomal history. All of the a few C/C individuals that we genotyped belong to the N2a1*-F1883(xF846) clade. On the other hand, there is no recurrent LLY22g C/C > C/A mutation event reported outside Haplogroup N. Moreover, an LLY22g A/A genotype inside N2b-F2569 clade also greatly favors the conversion theory. So we claim that the LLY22g C/C > C/A mutation is equivalent to M231, and downstream conversion events caused recurrent C/C or A/A alleles, and we cannot exclude the discovery of more recurrent events in Haplogroup N. The same case was also found for SNP P25 inside R1b clade, which has 3 copies on AZFc region in ampliconic region of Y chromosome [37]. Therefore, due to the prone-to-recurrence nature of LLY22g, we removed this SNP from the phylogeny of Haplogroup N and did no further genotyping of LLY22g in this study.

## Partial deletion in AZFc

There has been an observation that individuals of Haplogroup N had a b2/b3 deletion (about 1.8 Mb deleted, including locus sY1191) at the AZFc region of Y chromosome [38]. In that study, 31 of 33 males of Haplogroup N1 (i.e. M231+, LLY22g+) were found carrying b2/b3 deletion, and the other 2 suffered complete AZFc deletion and were infertile. As well, all the four N* (i.e. M231+, LLY22g-) males had b2/b3 deletion. In order to further confirm that whether all the lineages of Haplogroup N carry b2/b3 deletion, i.e. whether this deletion occurred already before the split of N1-F1206 and N2-F2930, we tested the existence of loci sY1191 (showing b2/b3 deletion) and

sY1291 (showing gr/gr deletion) on samples of most clades we identified in this paper. The result (Fig. 4) clearly showed that all the samples from both N1 and N2 subclades lacked sY1191 locus, indicating that b2/b3 deletion was already carried by the most recent common ancestor of N1-F1206 and N2-F2930, and should be inherited by all the individuals of Haplogroup N.

## Conclusion

Due to the fact that nearly all the samples north to China belong to a few downstream clades of N1-F1206, and China contains highest diversity of Haplogroup N and all the detected deep clades, as well as that Haplogroup N's closest relative – Haplogroup O inhabits also mainly in East and Southeast Asia, we suggest that Haplogroup N first diverged since the late Paleolithic Era inside the present range of China. The newly found SNPs through high-throughput sequencing become a powerful tool for better understanding of Haplogroup N, and will provide important information for early population history of Sino-Tibetan, Altaic and Uralic populations.

# Acknowledgement:

This study was supported by the National Science Foundation of China (NSFC) grant 31271338, 31071096, 31260252, 30760097, Research Foundation from Ministry of Education of China grant 311016, Social Science Foundation of Chinese Ministry of Education 12YJA850011, Key Project of the Natural Science Foundation of Tibet 201122 and the Natural Science Foundation of Tibet University for Nationalities (2013).

And my great thanks to arXiv, to let me avoid some reviewers that are so mean to the people they don't know well and simply deny the other researchers' hard work by "poor English" and delay the revision for the others till the deadline in order to let their own work publish first and say the others' work are not novel enough.

## Conflict of interest:

The authors declare no conflict of interest.

## References:

1. Underhill PA, Kivisild T: Use of Y chromosome and mitochondrial DNA population structure in tracing human migrations. *Annu Rev Genet* 2007; **41:** 539-564.

2. Cui Y, Li H, Ning C *et al*: Y Chromosome analysis of prehistoric human populations in the West Liao River Valley, Northeast China. *BMC Evol Biol* 2013; **13:** 216-216.

3. Rootsi S, Zhivotovsky LA, Baldovič M *et al*: A counter-clockwise northern route of the Y-chromosome haplogroup N from Southeast Asia towards Europe. *Eur J Hum Genet* 2007; **15:** 204-211.

4. Balanovsky O, Rootsi S, Pshenichnov A *et al*: Two sources of the Russian patrilineal heritage in their Eurasian context. *Am J Hum Genet* 2008; **82:** 236-250.

5. Pakendorf B, Novgorodov IN, Osakovskij VL, Danilova AP, Protod'jakonov AP, Stoneking M: Investigating the effects of prehistoric migrations in Siberia: genetic variation and the origins of Yakuts. *Hum Genet* 2006; **120:** 334-353.

6. Mirabal S, Regueiro M, Cadenas AM *et al*: Y-Chromosome distribution within the geo-linguistic landscape of northwestern Russia. *Eur J Hum Genet* 2009; **17:** 1260-1273.

7. Hammer MF, Karafet TM, Park H *et al*: Dual origins of the Japanese: common ground for hunter-gatherer and farmer Y chromosomes. *J Hum Genet* 2006; **51:** 47-58.

8. Zhong H, Shi H, Qi XB *et al*: Extended Y chromosome investigation suggests postglacial migrations of modern humans into East Asia via the northern route. *Mol Biol Evol* 2011; **28:** 717-727.

9. Tambets K, Rootsi S, Kivisild T *et al*: The western and eastern roots of the Saami - The story of genetic "outliers" told by mitochondrial DNA and Y chromosomes. *Am J Hum Genet* 2004; **74:** 661-682.

10. Derenko M, Malyarchuk B, Denisova G *et al*: Y-chromosome haplogroup N dispersals from south Siberia to Europe. *J Hum Genet* 2007; **52:** 763-770.

11. Kang L, Lu Y, Wang C *et al*: Y-chromosome O3 Haplogroup Diversity in Sino-Tibetan Populations Reveals Two Migration Routes into the Eastern Himalayas. *Ann Hum Genet* 2012; **76:** 92-99.

12. Kharkov VN, Stepanov VA, Medvedeva OF *et al*: The origin of Yakuts: Analysis of the Y-chromosome haplotypes. *Molecular Biology* 2008; **42:** 198-208.

13. Lappalainen T, Koivumaki S, Salmela E *et al*: Regional differences among the finns: A Y-chromosomal perspective. *Gene* 2006; **376:** 207-215.

14. Lappalainen T, Laitinen V, Salmela E *et al*: Migration waves to the Baltic Sea region. *Ann Hum Genet* 2008; **72:** 337-348.

15. Yan S, Wang CC, Li H, Li SL, Jin L, The Genographic Consortium: An updated tree of Y-chromosome Haplogroup O and revised phylogenetic positions of mutations P164 and PK4. *Eur J Hum Genet* 2011; **19:** 1013-1015.

16. Qi X, Cui C, Peng Y *et al*: Genetic Evidence of Paleolithic Colonization and Neolithic Expansion of Modern Humans on the Tibetan Plateau. *Mol Biol Evol* 2013; **30:** 1761-1778.

17. Karafet TM, Mendez FL, Meilerman MB, Underhill PA, Zegura SL, Hammer MF: New binary polymorphisms reshape and increase resolution of the human Y chromosomal haplogroup tree. *Genome Res* 2008; **18:** 830-838.

18. Yan S, Wang C-C, Zheng H-X *et al*: Y chromosomes of 40% chinese descend from three neolithic super-grandfathers. *PLoS One* 2014; **9**.

19. Inagaki S, Yamamoto Y, Doi Y *et al*: Typing of Y chromosome single nucleotide polymorphisms in a Japanese population by a multiplexed single nucleotide primer extension reaction. *Leg Med (Tokyo)* 2002; **4:** 202-206.

20. Repping S, Skaletsky H, Brown L *et al*: Polymorphism for a 1.6-Mb deletion of the human Y chromosome persists through balance between recurrent mutation and haploid selection. *Nat Genet* 2003; **35:** 247-251.

21. Firasat S, Khaliq S, Mohyuddin A *et al*: Y-chromosomal evidence for a limited Greek contribution to the Pathan population of Pakistan. *Eur J Hum Genet* 2007; **15:** 121-126.

22. Regueiro M, Cadenas AM, Gayden T, Underhill PA, Herrera RJ: Iran: Tricontinental nexus for Y-chromosome driven migration. *Human Heredity* 2006; **61:** 132-143.

23. Sengupta S, Zhivotovsky LA, King R *et al*: Polarity and temporality of high-resolution Y-chromosome distributions in India identify both indigenous and exogenous expansions and reveal minor genetic influence of central Asian pastoralists. *Am J Hum Genet* 2006; **78:** 202-221.

24. Cinnioğlu C, King R, Kivisild T *et al*: Excavating Y-chromosome haplotype strata in Anatolia. *Hum Genet* 2004; **114:** 127-148.

25. Xue YL, Zejal T, Bao WD *et al*: Male demography in East Asia: A north-south contrast in human population expansion times. *Genetics* 2006; **172:** 2431-2439.

26. Kharkov VN, Stepanov VA, Feshchenko SP, Borinskaya SA, Yankovsky NK, Puzyrev VP: Frequencies of Y chromosome binary haplogroups in Belarussians. *Russian Journal of Genetics* 2005; **41:** 928-931.

27. Kharkov VN, Stepanov VA, Medvedeva OF *et al*: Gene pool differences between Northern and Southern Altaians inferred from the data on Y-chromosomal haplogroups. *Russian Journal of Genetics* 2007; **43:** 551-562.

28. Pimenoff VN, Comas D, Palo JU, Vershubsky G, Kozlov A, Sajantila A: Northwest Siberian Khanty and Mansi in the junction of West and East Eurasian gene pools as revealed by uniparental markers. *Eur J Hum Genet* 2008; **16:** 1254-1264.

29. Puzyrev VP, Stepanov VA, Golubenko MV *et al*: MtDNA and Y-chromosome lineages in the Yakut population. *Russian Journal of Genetics* 2003; **39:** 816-822.

30. Zhivotovsky LA, Underhill PA, Cinnioğlu C *et al*: The effective mutation rate at Y chromosome short tandem repeats, with application to human population-divergence time. *Am J Hum Genet* 2004; **74:** 50-61.

31. Di Giacomo F, Luca F, Popa LO *et al*: Y chromosomal haplogroup J as a signature of the post-neolithic colonization of Europe. *Hum Genet* 2004; **115:** 357-371.

32. Slatkin M: A Measure of Population Subdivision Based on Microsatellite Allele Frequencies. *Genetics* 1995; **139:** 457-462.

33. Ge JY, Budowle B, Aranda XG, Planz JV, Eisenberg AJ, Chakraborty R: Mutation rates at Y chromosome short tandem repeats in Texas populations. *Forensic Sci Int-Gen* 2009; **3:** 179-184.

34. Goedbloed M, Vermeulen M, Fang RN *et al*: Comprehensive mutation analysis of 17 Y-chromosomal short tandem repeat polymorphisms included in the AmpFlSTR (R) Yfiler (R) PCR amplification kit. *Int J Legal Med* 2009; **123:** 471-482.

35. Skaletsky H, Kuroda-Kawaguchi T, Minx PJ *et al*: The male-specific region of the human Y chromosome is a mosaic of discrete sequence classes. *Nature* 2003; **423:** 825-U822.

36. Lange J, Skaletsky H, Bell GW, Page DC: MSY Breakpoint Mapper, a database of sequence-tagged sites useful in defining naturally occurring deletions in the human Y chromosome. *Nucleic Acids Res* 2008; **36:** D809-D814.

37. Adams SM, King TE, Bosch E, Jobling MA: The case of the unreliable SNP: Recurrent back-mutation of Y-chromosomal marker P25 through gene conversion. *Forensic Sci Int* 2006; **159:** 14-20.

38. Zhang F, Lu C, Li Z *et al*: Partial deletions are associated with an increased risk of complete deletion in AZFc: a new insight into the role of partial AZFc deletions in male infertility. *Journal of Medical Genetics* 2007; **44:** 437-444.

# Figure legend:

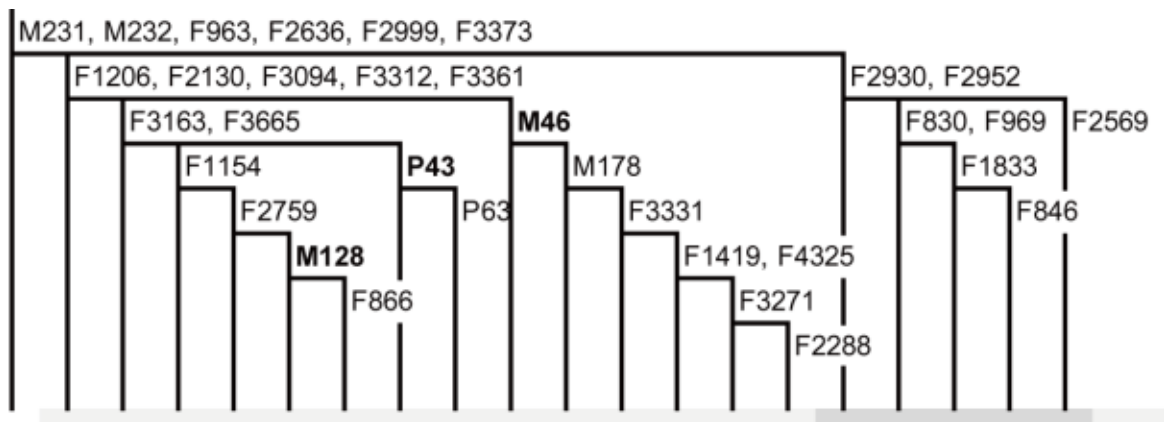

| Population | Sample type | Language family | Language clade | sum | Hap. N | not further geno-typed | further geno-typed | N* | N1* | N1a* | N1a1* | N1a1a* | N1a1a1* | N1a1a1a | N1a2* | N1a2a | N1b* | N1b1* | N1b1a* | N1b1a1* | N1b1a1a* | N1b1a1a1 | N2* | N2a* | N2a1* | N2a1a | N2b | N1 total | % | N2 total | % |
|---|---|---|---|---|---|---|---|---|---|---|---|---|---|---|---|---|---|---|---|---|---|---|---|---|---|---|---|---|---|---|---|
| Han, Northeast | Blood | Sino-Tibetan | Sinitic | 84 | 4 | 1 | 3 | 0 | 0 | 0 | 0 | 1 | 0 | 0 | 0 | 0 | 0 | 0 | 1 | N/A | 0 | 0 | 0 | 0 | 1 | 0 | 0 | 2 | 3.2 | 1 | 1.6 |
| Han, North | Blood | Sino-Tibetan | Sinitic | 219 | 12 | 0 | 12 | 0 | 1 | 0 | 0 | 2 | 1 | 0 | 0 | 0 | 0 | 0 | 1 | 2 | 0 | 1 | 0 | 0 | 2 | 0 | 2 | 8 | 3.7 | 4 | 1.8 |
| Han, Northwest | Blood | Sino-Tibetan | Sinitic | 44 | 4 | 0 | 4 | 0 | 0 | 0 | 0 | 0 | 0 | 0 | 0 | 0 | 0 | 1 | N/A | 0 | 0 | 0 | 0 | 0 | 2 | 0 | 1 | 1 | 2.3 | 3 | 6.8 |
| Han, East | Blood | Sino-Tibetan | Sinitic | 499 | 35 | 1 | 34 | 0 | 3 | 0 | 3 | N/A | 3 | 1 | 0 | 0 | 0 | 2 | 2 | 1 | 0 | 0 | 0 | 0 | 12 | 0 | 7 | 15 | 3.1 | 19 | 3.9 |
| Han, South | Blood | Sino-Tibetan | Sinitic | 173 | 10 | 0 | 10 | 0 | 0 | 0 | 0 | 0 | 2 | 0 | 0 | 0 | 0 | 1 | 1 | 0 | 0 | 0 | 0 | 0 | 3 | 1 | 2 | 4 | 2.3 | 6 | 3.5 |
| Han, Southwest | Blood | Sino-Tibetan | Sinitic | 34 | 1 | 0 | 1 | 0 | 0 | 0 | 0 | 0 | 0 | 0 | 0 | 0 | 0 | 0 | 0 | 0 | 0 | 0 | 0 | 0 | 1 | 0 | 0 | 0 | 0.0 | 1 | 2.9 |
| Han, Gansu | Saliva | Sino-Tibetan | Sinitic | 1126 | 91 | 14 | 77 | 0 | 11 | 0 | 0 | 11 | 3 | 0 | 0 | 0 | 0 | 8 | 1 | N/A | 0 | 1 | 5 | 3 | 28 | 1 | 5 | 35 | 3.7 | 42 | 4.4 |
| Han, Shandong | Saliva | Sino-Tibetan | Sinitic | 874 | 71 | 4 | 67 | 0 | 4 | 0 | 0 | 5 | 12 | 0 | 0 | 0 | 0 | 2 | 12 | N/A | 1 | 0 | 0 | 3 | 16 | 0 | 12 | 36 | 4.4 | 31 | 3.8 |
| Han, Sichuan, Suining | Saliva | Sino-Tibetan | Sinitic | 121 | 14 | 8 | 6 | 0 | 0 | 0 | 0 | 0 | 0 | 0 | 0 | 0 | 0 | 0 | 0 | 0 | 0 | 0 | 3 | 0 | 3 | 0 | 0 | 0 | 0.0 | 6 | 11.6 |
| Han, Heilongjiang, Wuchang | Saliva | Sino-Tibetan | Sinitic | 34 | 13 | 6 | 7 | 0 | 1 | 0 | 0 | 0 | 2 | 0 | 0 | 0 | 0 | 0 | 0 | 0 | 0 | 0 | 1 | 0 | 3 | 0 | 0 | 3 | 16.4 | 4 | 21.8 |
| Han, Guangxi, Pingguo | Saliva | Sino-Tibetan | Sinitic | 19 | 1 | 0 | 1 | 0 | 0 | 0 | 0 | 0 | 0 | 0 | 0 | 0 | 0 | 0 | 0 | 0 | 0 | 0 | 0 | 0 | 1 | 0 | 0 | 0 | 0.0 | 1 | 5.3 |
| Hui | Blood / Saliva | Sino-Tibetan | Sinitic | 286 | 15 | 2 | 13 | 0 | 0 | 0 | 0 | 0 | 1 | 0 | 0 | 0 | 0 | 2 | 1 | N/A | 1 | 0 | 2 | 0 | 5 | 0 | 1 | 5 | 2.0 | 8 | 3.2 |
| Mongol, Bargud, Hulun Buir | Blood | Altaic | Mongolic | 23 | 4 | 0 | 4 | 0 | 0 | 0 | 0 | 1 | 0 | 0 | 0 | 0 | 0 | 0 | 0 | 0 | 1 | 2 | 0 | 0 | 0 | 0 | 0 | 4 | 17.4 | 0 | 0.0 |
| Mongol, Buriyad, Hulun Buir | Blood | Altaic | Mongolic | 53 | 33 | 0 | 33 | 0 | 1 | 0 | 0 | 0 | 0 | 0 | 0 | 0 | 0 | 0 | 0 | 6 | 25 | 1 | 0 | 0 | 0 | 0 | 0 | 33 | 62.3 | 0 | 0.0 |
| Mongol, Ööled, Hulun Buir | Blood | Altaic | Mongolic | 62 | 3 | 0 | 3 | 0 | 0 | 0 | 1 | 0 | 0 | 0 | 1 | 0 | 0 | 1 | N/A | 0 | 0 | 0 | 0 | 0 | 0 | 0 | 0 | 3 | 4.8 | 0 | 0.0 |
| Mongol, others | Blood | Altaic | Mongolic | 48 | 3 | 0 | 3 | 0 | 1 | 0 | 0 | 0 | 0 | 0 | 1 | 0 | 0 | 0 | 0 | 0 | 1 | 0 | 0 | 0 | 0 | 0 | 0 | 3 | 6.3 | 0 | 0.0 |
| Tibetan, Ü-Tsang | Saliva | Sino-Tibetan | Bodic | 788 | 60 | 9 | 51 | 0 | 0 | 0 | 0 | 0 | 0 | 0 | 0 | 0 | 0 | 3 | N/A | N/A | 0 | 1 | 0 | 0 | 45 | 2 | 0 | 4 | 0.6 | 47 | 7.0 |
| Tibetan, Khams | Saliva | Sino-Tibetan | Bodic | 429 | 23 | 2 | 21 | 0 | 0 | 0 | 0 | 0 | 0 | 0 | 0 | 0 | 0 | 1 | 0 | 0 | 0 | 0 | 0 | 0 | 14 | 4 | 2 | 1 | 0.3 | 20 | 5.1 |
| Tibetan, Amdo | Saliva | Sino-Tibetan | Bodic | 506 | 19 | 4 | 15 | 0 | 0 | 0 | 0 | 0 | 0 | 1 | 0 | 0 | 0 | 0 | 0 | 0 | 0 | 0 | 0 | 0 | 11 | 1 | 2 | 1 | 0.3 | 14 | 3.5 |
| Tibetan, Xining | Saliva | Sino-Tibetan | Bodic | 122 | 5 | 3 | 2 | 0 | 0 | 0 | 0 | 0 | 0 | 0 | 0 | 0 | 0 | 0 | 0 | 0 | 0 | 0 | 1 | 0 | 1 | 0 | 0 | 0 | 0.0 | 2 | 4.1 |
| Tibetan, Dingri | Blood | Sino-Tibetan | Bodic | 113 | 7 | 1 | 6 | 0 | 0 | 0 | 0 | 0 | 0 | 0 | 0 | 0 | 0 | 0 | 0 | 0 | 0 | 0 | 0 | 0 | 2 | 0 | 4 | 0 | 0.0 | 6 | 6.2 |
| Tibetan, Gongbu | Blood | Sino-Tibetan | Bodic | 104 | 3 | 0 | 3 | 0 | 0 | 0 | 0 | 0 | 0 | 0 | 0 | 0 | 0 | 0 | 0 | 0 | 0 | 0 | 0 | 0 | 3 | 0 | 0 | 0 | 0.0 | 3 | 2.9 |
| Tibetan, Gannan | Blood | Sino-Tibetan | Bodic | 264 | 9 | 0 | 9 | 0 | 0 | 0 | 0 | 1 | 1 | 0 | 0 | 0 | 0 | 0 | 0 | 0 | 0 | 0 | 0 | 0 | 5 | 0 | 2 | 2 | 0.8 | 7 | 2.7 |
| Sherpa | Blood | Sino-Tibetan | Bodic | 87 | 0 | 0 | 0 | 0 | 0 | 0 | 0 | 0 | 0 | 0 | 0 | 0 | 0 | 0 | 0 | 0 | 0 | 0 | 0 | 0 | 0 | 0 | 0 | 0 | 0.0 | 0 | 0.0 |
| Mönpa | Saliva | Sino-Tibetan | Bodic | 15 | 1 | 0 | 1 | 0 | 0 | 0 | 0 | 0 | 0 | 0 | 0 | 0 | 0 | 0 | 0 | 0 | 0 | 0 | 0 | 0 | 1 | 0 | 0 | 0 | 0.0 | 1 | 6.7 |
| Mönpa | Blood | Sino-Tibetan | Bodic | 34 | 0 | 0 | 0 | 0 | 0 | 0 | 0 | 0 | 0 | 0 | 0 | 0 | 0 | 0 | 0 | 0 | 0 | 0 | 0 | 0 | 0 | 0 | 0 | 0 | 0.0 | 0 | 0.0 |
| Deng | Blood | Sino-Tibetan | Mishmi | 61 | 1 | 0 | 1 | 0 | 0 | 0 | 0 | 0 | 0 | 0 | 0 | 0 | 0 | 0 | 0 | 0 | 0 | 0 | 0 | 0 | 1 | 0 | 0 | 0 | 0.0 | 1 | 1.6 |
| Lhoba | Blood | Sino-Tibetan | Tani | 109 | 39 | 8 | 31 | 0 | 0 | 0 | 0 | 0 | 0 | 0 | 0 | 0 | 0 | 0 | 0 | 0 | 0 | 0 | 1 | 0 | 30 | 0 | 0 | 0 | 0.0 | 31 | 35.8 |
| Jiarong | Saliva | Sino-Tibetan | Jiarong | 94 | 4 | 2 | 2 | 0 | 0 | 0 | 0 | 0 | 0 | 0 | 0 | 0 | 0 | 0 | 0 | 0 | 0 | 0 | 0 | 0 | 1 | 1 | 0 | 0 | 0.0 | 2 | 4.3 |
| Queyu | Saliva | Sino-Tibetan | Qiangic | 135 | 5 | 1 | 4 | 0 | 0 | 0 | 0 | 0 | 0 | 0 | 0 | 0 | 0 | 1 | N/A | N/A | 0 | 0 | 0 | 0 | 3 | 0 | 0 | 1 | 0.9 | 3 | 2.8 |
| Muya | Saliva | Sino-Tibetan | Qiangic | 140 | 13 | 7 | 6 | 0 | 0 | 0 | 0 | 0 | 0 | 0 | 0 | 0 | 0 | 0 | 0 | 0 | 0 | 0 | 0 | 0 | 4 | 2 | 0 | 0 | 0.0 | 6 | 9.3 |
| Qiang | Blood | Sino-Tibetan | Qiangic | 149 | 6 | 0 | 6 | 0 | 0 | 0 | 0 | 1 | 0 | 0 | 0 | 0 | 0 | 0 | 0 | 0 | 0 | 0 | 0 | 0 | 2 | 2 | 1 | 1 | 0.7 | 5 | 3.4 |
| Tujia | Blood | Sino-Tibetan | Tujia | 42 | 1 | 0 | 1 | 0 | 0 | 0 | 0 | 0 | 0 | 0 | 0 | 0 | 0 | 0 | 0 | 0 | 0 | 0 | 0 | 0 | 1 | 0 | 0 | 0 | 0.0 | 1 | 2.4 |
| Zhuang, Guangxi, Pingguo | Saliva | Tai-Kadai | Daic | 29 | 3 | 0 | 3 | 0 | 0 | 0 | 0 | 0 | 0 | 0 | 0 | 0 | 0 | 0 | 0 | 0 | 0 | 0 | 0 | 0 | 3 | 0 | 0 | 0 | 0.0 | 3 | 10.3 |

Fig. 1 Refined phylogenetic tree of Haplogroup N, and the frequency of its subclades genotyped in populations in this study. Several samples collected several years ago that were genotyped as Haplogroup N but could not be further genotyped for subclades are labeled here as "not further genotyped", and the percentage of N1 results in the population (3rd last column) is calculated as N1/[further genotyped N]/([all N]/[all samples]) in order to avoid bias, likewise also for N2. The three SNPs in bold are previously well-studied as N1a-M128, N1b-P43, and N1c-M46.

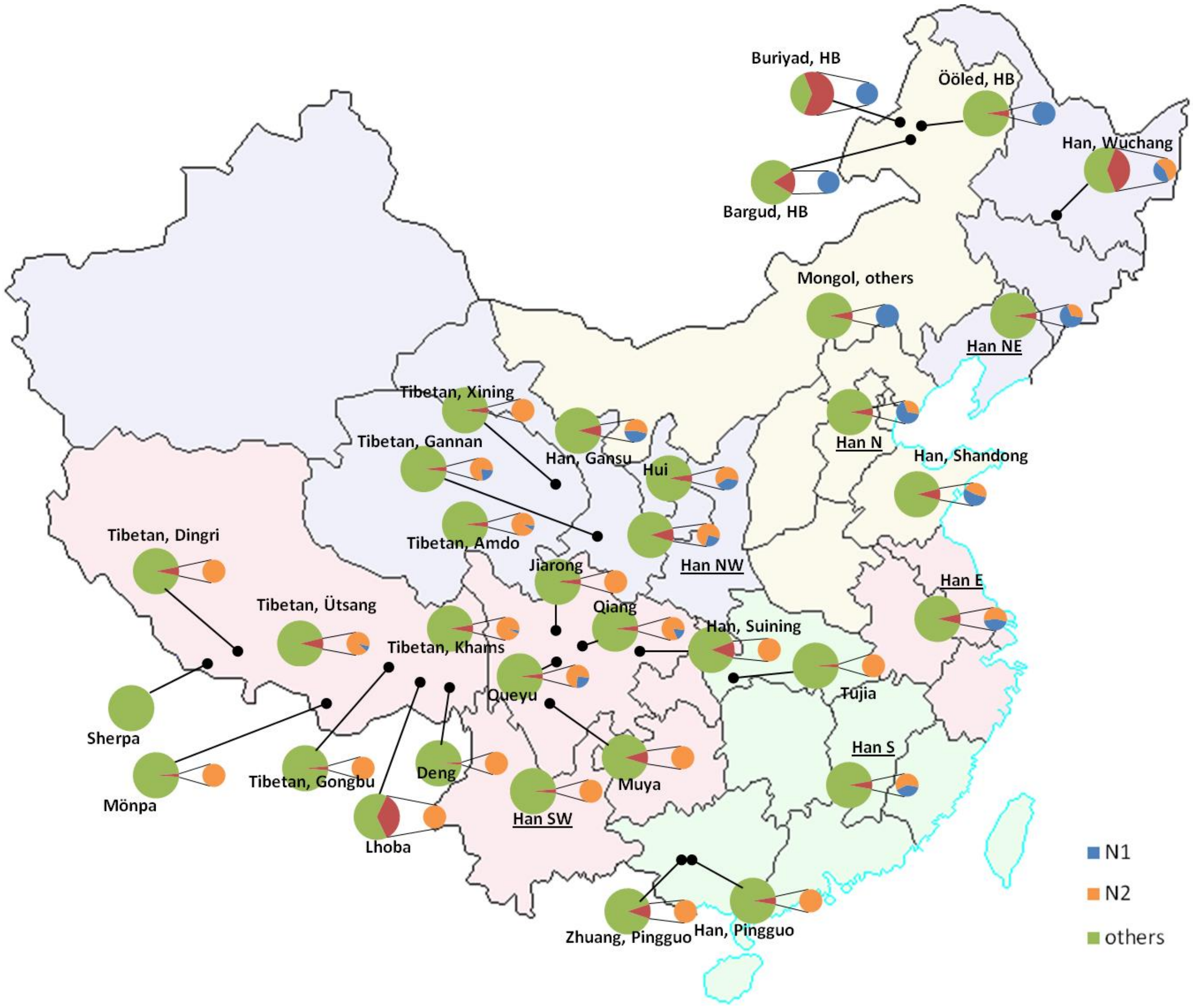

Fig. 2 Distribution of N1 and N2 genotyped in this study. For each population, the red part in the left circle represents the total proportion of N in the whole sample, while the right circle shows N1/N2 ratio. Han Chinese (labels with underline) were separated into six geographical parts with different colors in the map.

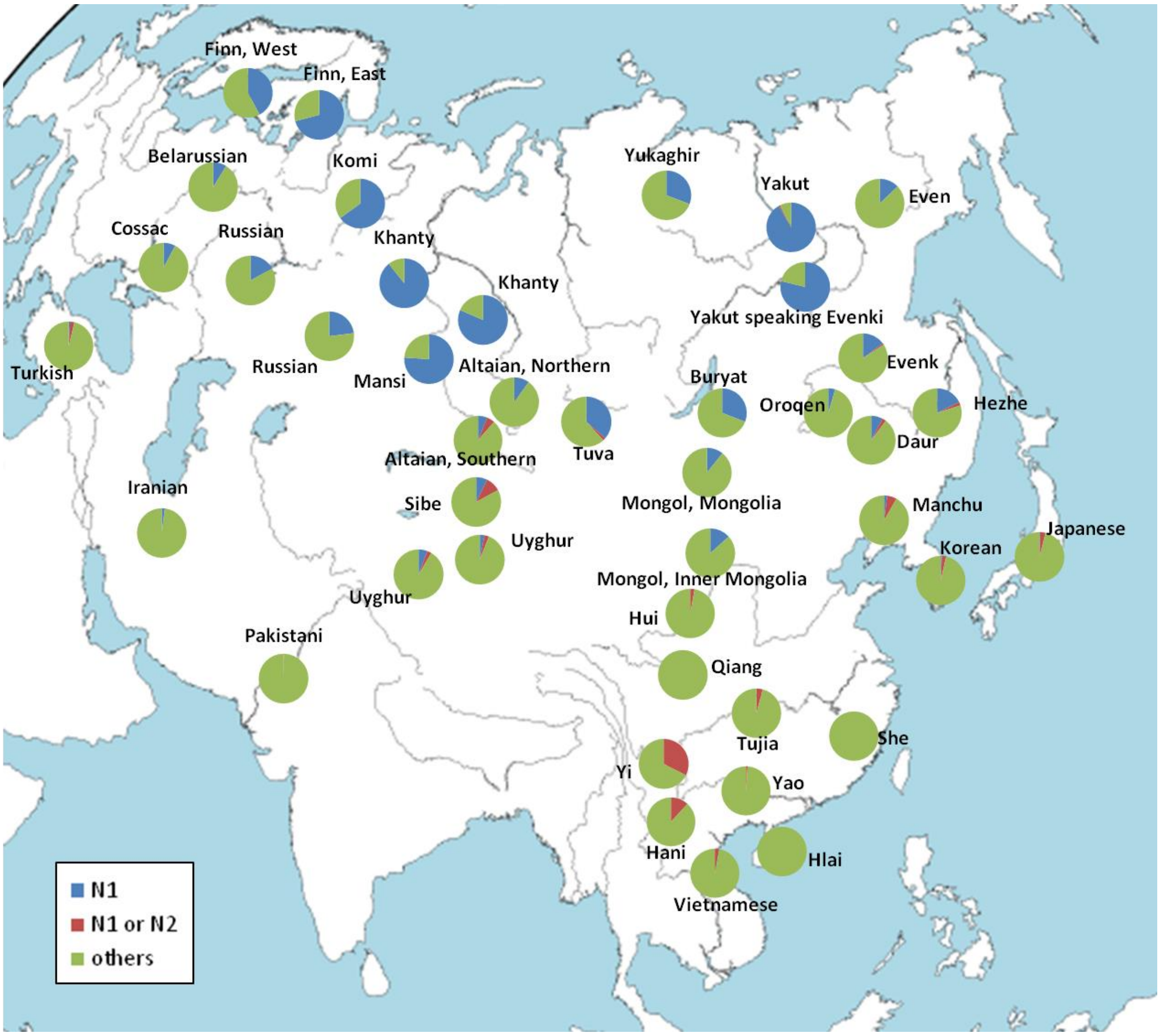

Fig. 3 Distribution of N1 (as the sum of N1a1a1-M128, N1a2-P43, and N1b-M46) and other N's of various groups from the literatures.

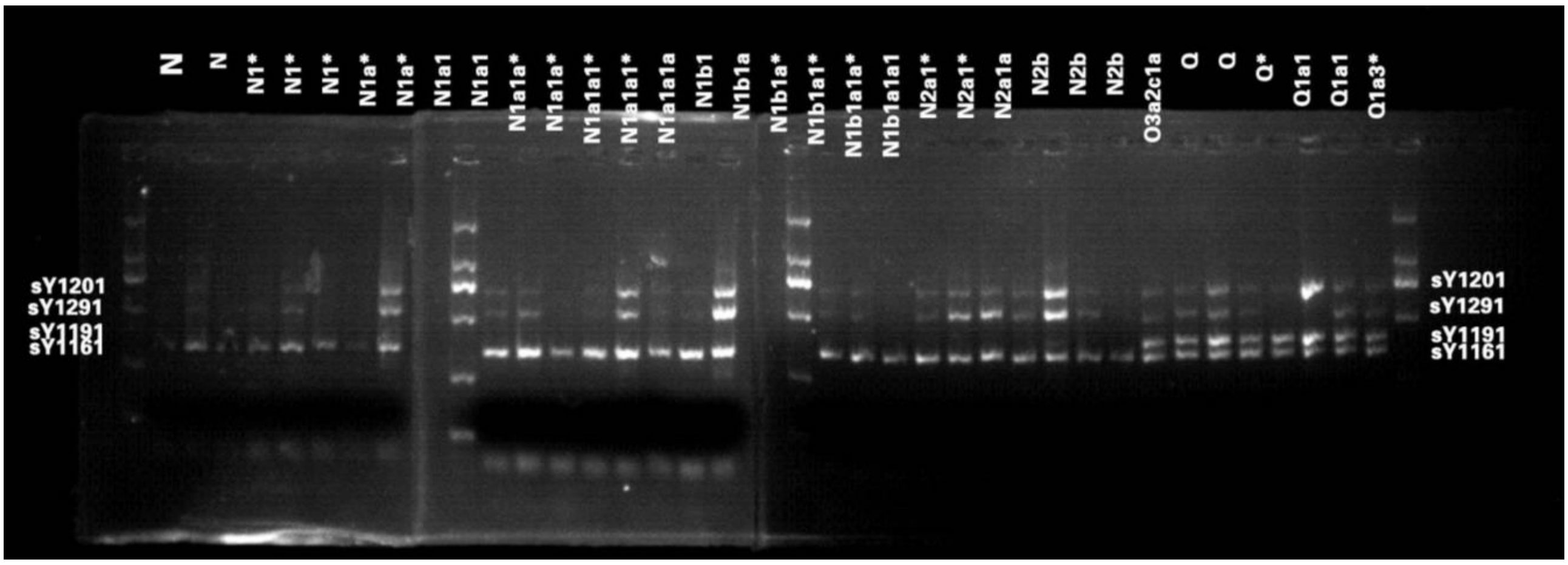

Fig. 4 Partial deletion b2/b3 in AZFc, showing non-existence of sY1191 band at samples from the all tested subclades of Haplogroup N, in contrast to samples of Haplogroup O and Q.

# Tables

Table 1 Calculated coalescence times of Haplogroup N and its subclades

| clade | population | coalescence time (kya) | linearly calibrated (kya) |
|---|---|---|---|
| N-M231 | mainly Han and Mongol | 9.8 ± 6.2 | 15.8 ± 10.0 |
| N1-F2130 | mainly Han and Mongol | 8.4 ± 5.0 | 13.5 ± 8.1 |
| N1a1-F1154 | mainly Han | 3.1 ± 4.1 | 5.0 ± 6.6 |
| N1b1-M178 | Han and Mongol | 7.8 ± 8.0 | 12.6 ± 12.9 |
| N1b1a1a-F3271 | mainly Mongol | 1.2 ± 1.5 | 1.9 ± 2.4 |
| N2-F2930 | mainly Han | 6.7 ± 4.4 | 10.8 ± 7.1 |
| N2a1-F1833 | mainly Han | 5.1 ± 4.2 | 9.2 ± 6.8 |
| N2b-F2569 | Han | 7.0 ± 3.0 | 11.2 ± 4.8 |

SI Table 1 Primers used in this study

| SNP | Clade | Position hg18 | Position hg19 | Forward primer (5′ > 3′) | Reverse primer (5′ > 3′) | Extension primer (5′ > 3′) | Mutation (hg direction) | Panel |
|---|---|---|---|---|---|---|---|---|
| P43 | N1a2 | 20340361 | 21880973 | TTCAGTTTAAAACATTAGTAGGCATACA | TCTTTTAACCGAAGGGCTATACTTCT | AGGCTGAGGCAGGAGAATGGC | G->A | 1 |
| F2130 | N1 | 14839392 | 16329998 | TAATAATTGCATACAGCTGAGGAATGG | TATAATCACTTTCATCTTCTCAATAATAAA | CATAGAGAATGGCAGTATATGGAT | G->A | 1 |
| F1833 | N2a1 | 13283378 | 14773984 | TTAAGGACAGGAATTCTATAGCTGAAG | ACGTCTTCCTCTCTTTGCTAACGG | CTCAGTCCCGACTGAAGCGTATTCATCAC | G->C | 1 |
| F846 | N2a1a | 6936175 | 6876175 | GCTGCTTCTCAGAGTTTGCTTGGA | CTGTTATACCACATGACCTATCTTAAA | AGAGGAGAGAAAGATGACTCACACTGGGTAT | G->A | 1 |
| F866 | N1a1a1a | 6983781 | 6923781 | TTTACCGTCTTAGACAGTTACTGCAT | AGATTTCTGATTACAAGACACTATTTG | GTTTATTTGGGTGTGCACATATGAAGGAGGAG CA | G->A | 1 |
| F3361 | N1 | 21269135 | 22859747 | GACCCTGGCACACACAAAGGCA | ACCTTCACTGGTAGCAGACTTTCAG | GCAGTGTTACAGCTGGCAGCCCTAGGGCACAC TGGAAGCAT | T->C | 1 |
| F2636 | N | 16429880 | 17920486 | AGTCCAGATTCACTGCCACTGAAG | CTTTAAACTCTGGCATCTAAGAAATAC | ATTGCAGTTTGCAAGCACTTCAGGGAAAACTG CTCT | G->A | 1 |
| F2952 | N2 | 17657875 | 19148481 | CATCTGATCATGCACATCTAACCATT | GAGGTTCAACCACACTTAATGGATG | GGAATGTGCTGAATACAGGAAAATAGTGTTCCA TCTGCT | G->A | 1 |
| F2930 | N2 | 17589996 | 19080602 | TTGAACATGAACGTGCATATCACCC | GACTGCTGATACAAGATGATTTAGAT | TTAGTATATTGTCTGGTATATGGTAGGGATATA ATTTTTTAGT | G->A | 1 |
| F3163 | N1a | 19691251 | 21231863 | AAAGTGCCAGTGCTTCTCCAACGG | TCTTAGAATTTAACCTACACCTTCTCC | ACGAAGGCAGATGGAGGGTGTGTAAGCAGCAGT GACGTGGGAGATA | G->A | 1 |
| F2288 | N1b1a1a1 | 15345333 | 16835939 | GGTTAGTTCGTGAATAATATTCCTAGA | GTAGGTTTCAATTCAAGATCAAATGC | AGATCTCTGTAAAAACATTTGTTTTTTTGGGTA | G->C | 1 |
| | | | | | | AAAATCCATTAAAAT | | |
| M231 | N | 13979118 | 15469724 | AATTTAGTTAATCATCATTCAATTAATACC | GGTGGCCAGAGTCTTTCACATCA | TGACACCACAGAAATTACAGGTATGAATTCTTTGACGATCTTTCCCCCAATT | G->A | 1 |
| M178 | N1b1 | 20201143 | 21741755 | GGACACAGGCAGAGACTCCGAAA | TCTGCTACTTTAATGAGAATGAATAGG | CTGTCCCTGAATGAAACCAAAGGCCATATATAGTTCTCCTGGCACACTAAGGAGCT | C->T | 1 |
| F4325 | N1b1a1 | 21384171 | 22974783 | TTAGGAGGCGGTTCTCTCATACCT | TGGATTATGTTACATGTAGTTAATCACT | AATGAAAAATACTTCTATTACATAC | A->G | 2 |
| F969 | N2a | 7518829 | 7458829 | GCTGCCAGAAGAAGAGATGACTTTG | CTTGAGCCAGACATACCCTGATAA | TTTCCAGACCCACAGAAACTGTGAAGTAATTA | T->C | 2 |
| F3665 | N1a | 22905198 | 24495810 | TCTTGGTAAATGCTACTCCAACTATC | CAGTGGATCATTTACAGAAGGCAATT | TGACACACAGTGTAAGCTAGAACCATGTTTCTA | T->G | 2 |
| F3373 | N | 21331156 | 22921768 | TGAACTAAGGCAAGTCATCTACTCAA | GAGGAAGTGTTGCATACATATCTTCT | GCCTGCCATTGTCTTCTACAGGCACCTCGTCCATC | G->A | 2 |
| M46/Tat | N1b | 13431977 | 14922583 | CTTTGTCATTGATATGAAATATTGCCAA | TTGTAAGCATAATTGAGAAGGTGCCG | TTAAAAAAATTATATGGACTCTGAGTGTAGACTTGTGAATTCA | T->C | 2 |
| F2759 | N1a1a | 16925271 | 18415877 | GGAGGCACCTACTTAAAAACATGCC | TGTCATTGTGCATGGGAAAACTCCC | AATTCTGTAAATGCTTACCAGTGGCCTTCAGTGAGCACCCTCAC | A->G | 2 |
| F1206 | N1 | 8500417 | 8440417 | TGAAGACAGATGACACAGAAGGTGC | CCAGGTATGGCAACGTTCAAATAGG | GGTGCTTTCAACTCCATCCCTGAATTCCCTTAATTGCACAAGCAGTC | C->T | 2 |
| F830 | N2a | 6861568 | 6801568 | TGCAGGCTCTAAATGTGAAATCCCA | TGGAAAGCTGAGGGCACACACC | GGAGCTCATTGTTTTTGTGTTAACAGTAGGAAATGCTGCATAAAGAAGG | T->C | 2 |
| F1154 | N1a1 | 8441313 | 8381313 | AACTCATAAATATCTGGATGAAATGCTT | GACACTGTTTCAAAGTGCAGAGACT | AAATATCTGGATGAAATGCTTATAAAGAGGTCAGCCTTAAAAAATGTCATGAGG | T->C | 2 |
| F3271 | N1b1a1a | 20331453 | 21872065 | CCAAGAGCCAGTGTCTTATGTAGG | AACCTTCTCATAATAACAGGTTCTTTC | TATGTGTTCCCATACCATCTCTGTGTGTACAGCATGCAAGGAAATATCCTTCCAGAC | A->C | 2 |
| F1419 | N1b1a1 | 8864414 | 8804414 | GTTGGACAGGAAGGAAGTTCTTAAC | CCTTTTCCCAATATCTTCCATTCATG | CACATCTATCATTGATAGAAA | C->A | 3 |
| F3094 | N1 | 19286846 | 20827458 | CCTTCATGAGTGAATCCTAATCAACT | GAACAAGTCTCTCTACTTCAGTGTC | CCCAGATTCATAGCAGACATAGATAATTG | T->A | 3 |
| M128 | N1a1a1 | 20227318 | 21767930 | AATATATTTCTTGATGATGAGACCAATAT | GAACTGCCTCTTATAAAATCATATCCT | TCTACCTCTTTCAAACTGTAAATGAAAATAACTG | TG del | 3 |
| F963 | N | 7475220 | 7415220 | TGCAGTAACATCAATGGTAGGACATT | GTGGAAACTTAAGTTCATATCCTTGC | CAACTGGATTCATACAAATGGCCAGATGGCAAAGA | G->C | 3 |
| F2569 | N2b | 16212992 | 17703598 | GATCTTAAGACAGCAATTTCATTTCC | AGGGACAGAGGATTTCCTTCCTTC | TTTCAATGACTCCCCAGAATCAAGATGGAATACCCAT | A->G | 3 |
| P63 | N1a2a | 20303215 | 21843827 | CCACGGTCCTTGAAGCTTTCTGG | AACTTTCCATGAGAAGTGCTGTAGG | CCCAGGTCAGCACTTGCCCATCACTCTTCTAGGTATCAT | C ins | 3 |
| M232 | N | 13946546 | 15437152 | CCACAACAGGCAAACCTTAAATTTGT | CTCGAAACAGATCTGTACGAGAGG | GAAACGCTTATTTTTAGTCTCTCTTCCATGACTCTTCTAATA | C->T | 3 |
| F3331 | N1b1a | 21043894 | 22634506 | GGCTGTCTATAGTTGAGTCAACAATA | TGTTACATCAATTCAGTTACTCCTATAA | ACAAGAAGAAAGCCCTATGTAGTGCCTGCCCAAGGGGACTGAGTA | G->T | 3 |
| F2999 | N | 17778700 | 19269306 | AAATAATATTTCCTGGAGATTTCCAAAG | GTGGAAAGTTAGAATTTACATTCTGAC | GGAGATTTCCAAAGAAATATTTGGCAGCATAAACTAAACAGTGATTAGGAATC | C->T | 3 |
| F3312 | N1 | 20672123 | 22212735 | ACAATGGATAGATTTATCAACATGCATT | TATTAGGCCAAGTTCAAGTATTATACAG | AATGGATAGATTTATCAACATGCATTTTCACCCTCAGTTCACTGTGGTTCTGTTGTC | A->G | 3 |

SI Table 2 Frequency of subclades of Haplogroup N from literatures

| Population | Linguistic taxonomy | Reference | Sample size | N or NOxO | N% | N1a1a1-M128 | N1a2-P43 | N1b-M46, M178 | NO*(xM175,M128,P43,M46) |
|---|---|---|---|---|---|---|---|---|---|

| | | | | total | | | | | |
|---|---|---|---|---|---|---|---|---|---|
| Pakistani | Indo-European, Indo-Iranian | Fisarat S (2007) | 638 | 2 | 0% | N/A | N/A | 0 | 2 |
| Iranian | Indo-European, Indo-Iranian | Regueiro M (2006) | 150 | 3 | 2% | 0 | 3 | 0 | 0 |
| Russian | Indo-European, Slavic | Mirabal S (2009) | 106 | 18 | 17% | 0 | 0 | 18 | 0 |
| Russian | Indo-European, Slavic | Balanovski O (2008) | 1138 | 255 | 22% | 0 | 34 | 227 | 1 |
| Cossac | Indo-European, Slavic | Balanovski O (2008) | 90 | 7 | 8% | 0 | 1 | 6 | 0 |
| Belarussian | Indo-European, Slavic | Kharkov VN (2005) | 68 | 6 | 9% | 0 | 0 | 6 | 0 |
| Finn, East | Uralic | Lappalainen P (2006) | 306 | 217 | 71% | 0 | 0 | 217 | 0 |
| Finn, West | Uralic | Lappalainen P (2006) | 230 | 97 | 42% | 0 | 2 | 95 | 0 |
| Khanty | Uralic | Pimenoff VN (2008) | 28 | 25 | 89% | 0 | 7 | 18 | 0 |
| Khanty | Uralic | Mirabal S (2009) | 27 | 22 | 81% | 0 | 21 | 1 | 0 |
| Mansi | Uralic | Pimenoff VN (2008) | 25 | 19 | 76% | 0 | 15 | 4 | 0 |
| Komi | Uralic | Mirabal S (2009) | 103 | 67 | 65% | 0 | 16 | 51 | 0 |
| Yukaghir | Uralic | Pakendorf B (2006) | 13 | 4 | 31% | 0 | 0 | 4 | 0 |
| Yakut | Altaic, Turkic | Sengupta S (2006) | 18 | 16 | 89% | 0 | 0 | 16 | 0 |
| Yakut | Altaic, Turkic | Kharkov VN (2008) | 109 | 101 | 93% | 0 | 3 | 97 | 1 |
| Yakut | Altaic, Turkic | Pakendorf B (2006) | 184 | 174 | 95% | 0 | 1 | 173 | 0 |
| Yakut | Altaic, Turkic | Puzyrev VP (2003) | 46 | 42 | 91% | N/A | N/A | 40 | 2 |
| Yakut speaking Evenki | Altaic, Turkic | Pakendorf B (2006) | 33 | 26 | 79% | 0 | 2 | 24 | 0 |
| Tuva | Altaic, Turkic | Pakendorf B (2006) | 55 | 21 | 38% | N/A | 15 | 5 | 1 |
| Turkish | Altaic, Turkic | Cinnioğlu C (2004) | 523 | 20 | 4% | N/A | N/A | 5 | 15 |
| Uyghur | Altaic, Turkic | Xue YL (2006) | 70 | 6 | 9% | 0 | 2 | 2 | 2 |
| Uyghur | Altaic, Turkic | Hammer F (2006) | 67 | 4 | 6% | 0 | 2 | 0 | 2 |
| Altai | Altaic, Turkic | Hammer F (2006) | 98 | 4 | 4% | 0 | 2 | 1 | 1 |
| Altaian, Northern | Altaic, Turkic | Kharkov VN (2007) | 50 | 5 | 10% | 0 | 2 | 3 | 0 |
| Altaian, Southern | Altaic, Turkic | Kharkov VN (2007) | 96 | 11 | 11% | 0 | 4 | 2 | 5 |
| Daur | Altaic, Mongolic | Xue YL (2006) | 39 | 4 | 10% | 0 | 0 | 3 | 1 |
| Mongol, Inner Mongolia | Altaic, Mongolic | Xue YL (2006) | 45 | 6 | 13% | 0 | 0 | 6 | 0 |
| Mongol, Mongolia | Altaic, Mongolic | Xue YL (2006) | 65 | 7 | 11% | 0 | 2 | 5 | 0 |
| Buryat | Altaic, Mongolic | Hammer F (2006) | 81 | 25 | 31% | 0 | 2 | 23 | 0 |
| Mongol | Altaic, Mongolic | Hammer F (2006) | 149 | 13 | 9% | 0 | 9 | 3 | 1 |
| Even | Altaic, Tungusic | Hammer F (2006) | 31 | 4 | 13% | 0 | 0 | 4 | 0 |
| Evenk, Western | Altaic, Tungusic | Pakendorf B (2006) | 40 | 11 | 28% | 0 | 11 | 0 | 0 |
| Evenk, China | Altaic, Tungusic | Xue YL (2006) | 26 | 1 | 4% | 0 | 0 | 0 | 1 |
| Evenk, China | Altaic, Tungusic | Hammer F (2006) | 41 | 2 | 5% | 0 | 1 | 0 | 1 |
| Evenk | Altaic, Tungusic | Hammer F (2006) | 95 | 18 | 19% | 0 | 2 | 16 | 0 |
| Hezhe | Altaic, Tungusic | Xue YL (2006) | 45 | 9 | 20% | 0 | 8 | 0 | 1 |
| Manchu | Altaic, Tungusic | Xue YL (2006) | 35 | 5 | 14% | 2 | 1 | 0 | 2 |
| Manchu | Altaic, Tungusic | Hammer F (2006) | 96 | 8 | 8% | 2 | 0 | 0 | 6 |
| Oroqen | Altaic, Tungusic | Xue YL (2006) | 31 | 2 | 6% | 0 | 2 | 0 | 0 |
| Sibe | Altaic, Tungusic | Xue YL (2006) | 41 | 7 | 17% | 1 | 0 | 2 | 4 |
| Oroqen | Altaic, Tungusic | Hammer F (2006) | 22 | 1 | 5% | 0 | 0 | 1 | 0 |
| Japanese | Japanese | Xue YL | 47 | 3 | 6% | 0 | 0 | 0 | 3 |
| | | (2006) | | | | | | | |
| Japanese | Japanese | Hammer F (2006) | 255 | 10 | 4% | 0 | 0 | 1 | 9 |
| Korean, China | Korean | Xue YL (2006) | 25 | 1 | 4% | 0 | 0 | 0 | 1 |
| Korean, Korea | Korean | Xue YL (2006) | 43 | 2 | 5% | 0 | 0 | 0 | 2 |
| Korean | Korean | Hammer F (2006) | 75 | 2 | 3% | 1 | 0 | 0 | 1 |
| Hani | Sino-Tibetan, Lolo-Burmese | Xue YL (2006) | 34 | 4 | 12% | 0 | 0 | 0 | 4 |
| Yi | Sino-Tibetan, Lolo-Burmese | Hammer F (2006) | 43 | 14 | 33% | 0 | 0 | 0 | 14 |
| Qiang | Sino-Tibetan, Qiangic | Xue YL (2006) | 33 | 0 | 0% | 0 | 0 | 0 | 0 |
| Hui | Sino-Tibetan, Sinitic | Xue YL (2006) | 35 | 1 | 3% | 0 | 0 | 0 | 1 |
| Tujia | Sino-Tibetan, Tujia | Hammer F (2006) | 49 | 2 | 4% | 0 | 0 | 0 | 2 |
| Vietnamese | Austroasiatic, Vietic | Hammer F (2006) | 70 | 2 | 3% | 0 | 0 | 0 | 2 |
| Yao | Hmong-Mien | Hammer F (2006) | 60 | 0 | 0% | 0 | 0 | 0 | 0 |
| Yao, Bama | Hmong-Mien, Bunu | Xue YL (2006) | 35 | 1 | 3% | 0 | 0 | 0 | 1 |
| Miao | Hmong-Mien, Hmong | Hammer F (2006) | 58 | 0 | 0% | 0 | 0 | 0 | 0 |
| Yao, Liannan | Hmong-Mien, Mien | Xue YL (2006) | 35 | 0 | 0% | 0 | 0 | 0 | 0 |
| She | Hmong-Mien, She | Xue YL (2006) | 34 | 0 | 0% | 0 | 0 | 0 | 0 |
| She | Hmong-Mien, She | Hammer F (2006) | 51 | 0 | 0% | 0 | 0 | 0 | 0 |
| Hlai | Tai-Kadai, Hlai | Xue YL (2006) | 34 | 0 | 0% | 0 | 0 | 0 | 0 |
| Bouyei | Tai-Kadai, Tai | Xue YL (2006) | 35 | 4 | 11% | 2 | 0 | 0 | 2 |

SI Table 3 17-STR data of Haplogroup N

*Abbreviations: S = South, E = East, N = North, SW = Southwest, NE = Northeast, NW = Northwest, MN = Mongol, HB = Hulun Buir, Ö = Ööled, BR = Buriyad, BG = Bargud)

| Sample | Country | Ethnic group * | Haplogroup | Terminal SNPs | DYS-19 | DYS-389I | DYS-389b | DYS-390 | DYS-391 | DYS-392 | DYS-393 | DYS-437 | DYS-438 | DYS-439 | DYS-448 | DYS-456 | DYS-458 | DYS-635 | GATA-H4 | DYS-385a | DYS-385b |
|---|---|---|---|---|---|---|---|---|---|---|---|---|---|---|---|---|---|---|---|---|---|
| YCH3 | China | Han S | N2a1a | F846+ | 15 | 14 | 17 | 23 | 10 | 14 | 13 | 14 | 11 | 12 | 19 | 17 | 15 | 23 | 12 | 11 | 12 |
| YCH40 | China | Han E | N1a1a1a | F866+ | 14 | 13 | 16 | 23 | 10 | 16 | 13 | 14 | 10 | 11 | 19 | 15 | 17 | 21 | 12 | 13 | 13 |
| YCH43 | China | Han N | N1b1a* | F3331+, F1419-, F4325- | 14 | 14 | 15 | 23 | 10 | 14 | 13 | 14 | 10 | 11 | 18 | 15 | 19 | 20 | 11 | 12 | 12 |
| YCH74 | China | Han S | N1b1a* | F3331+, F1419-, F4325- | 14 | 14 | 16 | 24 | 11 | 14 | 13 | 16 | 10 | 10 | 18 | 14 | 17 | 20 | 11 | 11 | 11 |
| YCH107 | China | Han N | N2a1* | F1833+, F846- | 15 | 13 | 16 | 23 | 10 | 14 | 14 | 14 | 10 | 11 | 19 | 16 | 12 | 23 | 12 | 11 | 12 |
| YCH132 | China | Han S | N2a1* | F1833+, F846- | 14 | 14 | 16 | 23 | 11 | 14 | 13 | 14 | 11 | 11 | 19 | 16 | 16 | 23 | 11 | 12 | 12 |
| YCH133 | China | Han S | N2b | F2569+ | 14 | 14 | 18 | 22 | 11 | 14 | 13 | 14 | 10 | 11 | 19 | 16 | 18 | 21 | 11 | 11 | 13 |
| YCH141 | China | Han N | N1a1a1* | M128+, F866- | 14 | 13 | 16 | 23 | 11 | 15 | 13 | 14 | 10 | 11 | 19 | 15 | 17 | 21 | 12 | 12 | 13 |
| YCH142 | China | Han S | N2a1* | F1833+, F846- | 14 | 13 | 17 | 23 | 11 | 14 | 13 | 14 | 11 | 12 | 19 | 15 | 16 | 23 | 12 | 11 | 12 |
| YCH171 | China | Han E | N2b | F2569+ | 14 | 13 | 16 | 22 | 10 | 14 | 13 | 14 | 10 | 10 | 19 | 16 | 17 | 21 | 12 | 10 | 12 |
| YCH205 | China | Han N | N1* | F2130+, F3163-, M46- | 14 | 13 | 15 | 24 | 10 | 15 | 13 | 14 | 10 | 11 | 19 | 16 | 15 | 21 | 13 | 11 | 11 |
| YCH216 | China | Han SW | N2a1* | F1833+, F846- | 15 | 13 | 18 | 24 | 11 | 14 | 13 | 14 | 11 | 11 | 19 | 16 | 16 | 23 | 12 | 11 | 11 |
| YCH249 | Japan | Japanese | N1a1a1* | M128+, F866- | 14 | 13 | 16 | 22 | 10 | 15 | 13 | 14 | 11 | 11 | 19 | 15 | 16 | 21 | 12 | 12 | 13 |
| YCH287 | China | Han E | N2a1* | F1833+, F846- | 14 | 13 | 17 | 23 | 11 | 14 | 13 | 14 | 11 | 11 | 19 | 16 | 15 | 21 | 12 | 11 | 11 |
| YCH369 | China | Han N | N1a1a* | F2759+, M128- | 14 | 13 | 16 | 24 | 10 | 15 | 13 | 14 | 10 | 11 | 19 | 15 | 18 | 21 | 12 | 12 | 13 |
| YCH394 | China | Han N | N1b1a1a1 | F2288+ | 14 | 14 | 16 | 23 | 10 | 14 | 13 | 14 | 10 | 10 | 19 | 14 | 17 | 22 | 12 | 11 | 13 |
| YCH401 | China | Han E | N1b1a* | F3331+, F4325- | 14 | 14 | 16 | 23 | 10 | 14 | 14 | 14 | 10 | 10 | 18 | 14 | 16 | 20 | 11 | 11 | 12 |
| YCH403 | China | Han E | N2a1* | F1833+, F846- | 14 | 14 | 17 | 23 | 10 | 14 | 13 | 14 | 10 | 11 | 19 | 17 | 15 | 22 | 12 | 11 | 12 |
| YCH405 | China | Han N | N2a1* | F1833+, F846- | 14 | 14 | 16 | 23 | 11 | 14 | 13 | 14 | 11 | 13 | 19 | 15 | 15 | 23 | 12 | 11 | 12 |
| YCH406 | China | Han E | N2a1* | F1833+, F846- | 14 | 14 | 17 | 23 | 10 | 14 | 13 | 14 | 10 | 11 | 19 | 15 | 15 | 22 | 12 | 11 | 12 |
| YCH414 | China | Han E | N2a1* | F1833+, F846- | 14 | 14 | 17 | 23 | 10 | 14 | 13 | 14 | 10 | 11 | 19 | 17 | 16 | 24 | 12 | 11 | 12 |
| YCH431 | China | Han N | N1a1a* | F2759+, M128- | 15 | 13 | 16 | 24 | 10 | 15 | 13 | 14 | 10 | 11 | 19 | 15 | 16 | 21 | 12 | 12 | 13 |
| YCH433 | China | Han S | N1a1a1* | M128+, F866- | 14 | 13 | 17 | 23 | 10 | 15 | 13 | 14 | 10 | 11 | 20 | 15 | 16 | 22 | 12 | 12 | 13 |
| YCH437 | China | Han NW | N2a1* | F1833+, F846- | 14 | 14 | 15 | 23 | 10 | 14 | 14 | 14 | 11 | 11 | 19 | 16 | 17 | 22 | 12 | 11 | 12 |
| YCH481 | China | Han NE | N2a1* | F1833+, F846- | 14 | 13 | 16 | 23 | 10 | 13 | 13 | 14 | 10 | 12 | 20 | 16 | 16 | 24 | 12 | 11 | 12 |
| YCH529 | China | MN, Chahar | N1b1a1a* | F3271+, F2288- | 13 | 14 | 16 | 24 | 11 | 14 | 14 | 14 | 10 | 10 | 19 | 15 | 17 | 22 | 12 | 11 | 13 |
| YCH550 | China | Han NE | N1b1a | F3331+, F1419?, F4325?, F3271- | 14 | 14 | 16 | 24 | 11 | 14 | 13 | 14 | 10 | 10 | 18 | 14 | 21 | 20 | 12 | 10 | 13 |
| YCH554 | China | Han E | N1a1 | F2930+, F969-, F2569- | 14 | 13 | 16 | 23 | 10 | 15 | 13 | 15 | 10 | 12 | 19 | 15 | 16 | 21 | 12 | 12 | 13 |
| YCH566 | China | Han NW | N2b | F2569+ | 15 | 13 | 18 | 22 | 10 | 14 | 15 | 14 | 10 | 10 | 19 | 15 | 14 | 21 | 10 | 11 | 12 |
| YCH591 | China | MN | N1a2* | P43+, P63- | 14 | 13 | 15 | 23 | 10 | 14 | 13 | 14 | 10 | 10 | 19 | 15 | 16 | 23 | 12 | 12 | 13 |
| YCH595 | China | MN, Ö, Xinjiang | N1a2* | P43+, P63- | 14 | 14 | 16 | 23 | 10 | 14 | 13 | 14 | 10 | 10 | 19 | 15 | 16 | 24 | 12 | 12 | 13 |
| YCH630 | China | Han E | N2b | F2569+ | 15 | 13 | 15 | 22 | 10 | 15 | 13 | 14 | 10 | 11 | 19 | 16 | 16 | 23 | 12 | 11 | 14 |
| YCH633 | China | Han E | N2b | F2569+ | 15 | 13 | 15 | 22 | 10 | 15 | 13 | 14 | 10 | 11 | 19 | 16 | 16 | 23 | 12 | 11 | 15 |
| YCH641 | China | Han S | N2b | F2569+ | 14 | 13 | 17 | 22 | 10 | 14 | 13 | 14 | 10 | 11 | 20 | 15 | 19 | 22 | 12 | 11 | 15 |
| YCH644 | China | Han N | N1b1a1* | F1419+, F4325+, F3271- | 14 | 14 | 16 | 23 | 11 | 16 | 14 | 14 | 11 | 10 | 19 | 14 | 15 | 22 | 12 | 11 | 14 |
| YCH704 | China | Han S | N2a1* | F1833+, F846- | 15 | 15 | 18 | 23 | 11 | 14 | 13 | 14 | 11 | 12 | 19 | 15 | 16 | 23 | 12 | 11 | 12 |
| YCH732 | China | Han E | N1a1 | F1154+ | 14 | 14 | 16 | 23 | 10 | 15 | 13 | 14 | 10 | 11 | 19 | 15 | 16 | 21 | 13 | 12 | 13 |
| YCH751 | China | Yi, Guizhou | N2a1* | F1833+, F846- | 14 | 13 | 17 | 23 | 10 | 14 | 13 | 14 | 11 | 11 | 19 | 15 | 16 | 21 | 12 | 12 | 13 |
| YCH753 | China | Han E | N1a1 | F1154+ | 14 | 13 | 16 | 23 | 10 | 15 | 13 | 14 | 10 | 11 | 19 | 15 | 16 | 20 | 12 | 12 | 14 |
| YCH757 | China | Han E | N2b | F2569+ | 15 | 13 | 16 | 22 | 10 | 15 | 13 | 14 | 10 | 11 | 19 | 16 | 18 | 22 | 12 | 11 | 14 |
| YCH777 | China | Han NW | N2a1* | F1833+, F846- | 15 | 13 | 16 | 23 | 10 | 15 | 13 | 14 | 11 | 11 | 19 | 16 | 15 | 25 | 12 | 11 | 11 |
| YCH780 | China | Han NW | N1b1 | M178+, F3331?, F1419-, F4325- | 14 | 14 | 16 | 23 | 10 | 14 | 13 | 14 | 10 | 10 | 18 | 14 | 17 | 20 | 11 | 11 | 12 |
| YCH784 | China | Han N | N2b | F2569+ | 15 | 13 | 18 | 22 | 10 | 14 | 14 | 14 | 10 | 11 | 19 | 15 | ? | 22 | 11 | 11 | 14 |
| YCH786 | China | Han E | N2a1* | F1833+, F846- | 14 | 14 | 16 | 23 | 10 | 14 | 13 | 14 | 11 | 11 | 19 | 17 | 17 | 22 | 13 | 11 | 12 |
| YCH822 | China | Han NE | N1a1a1* | M128+, F710+, F3201+, F3275+, F866- | 14 | 13 | 16 | 23 | 10 | 15 | 13 | 14 | 10 | 11 | 19 | 15 | 16 | 22 | 12 | 12 | 13 |
| YCH824 | China | Han E | N1b1a1* | F1419+, F4325+, F3271- | 14 | 14 | 16 | 23 | 11 | 15 | 13 | 14 | 11 | 11 | 19 | 14 | 17 | 24 | 12 | 11 | 13 |
| YCH845 | China | Han N | N2b | F2569+ | 14 | 13 | 17 | 22 | 10 | 14 | 13 | 14 | 10 | 10 | 19 | 14 | ? | 22 | 11 | 11 | 12 |
| YCH848 | China | Han S | N1b1 | M178+, F3331?, F1419-, F4325- | 15 | 14 | 16 | 23 | 10 | 14 | 13 | 14 | 10 | 11 | 18 | 14 | 18 | 20 | 11 | 11 | 12 |
| YCH864 | China | Han E | N1a1 | F1154+ | 14 | 13 | 16 | 23 | 10 | 15 | 13 | 14 | 10 | 12 | 19 | 15 | 18 | 21 | 12 | 11 | 13 |
| YCH876 | China | Han E | N2b | F2569+ | 15 | 14 | 17 | 22 | 10 | 14 | 13 | 14 | 8 | 12 | 19 | 14 | 19 | 21 | 12 | 11 | 13 |
| YCH887 | China | Han E | N2b | F2569+ | 14 | 13 | 16 | 22 | 10 | 15 | 13 | 14 | 10 | 11 | 19 | 16 | 16 | 21 | 12 | 11 | 14 |
| YCH902 | China | Han E | N2a1* | F1833+, F846- | 14 | 14 | 17 | 23 | 10 | 14 | 13 | 14 | 10 | 11 | 19 | 18 | 15 | 23 | 12 | 11 | 12 |
| YCH909 | China | Han E | N2a1* | F1833+, F846- | 14 | 14 | 17 | 23 | 10 | 14 | 13 | 14 | 10 | 11 | 19 | 17 | 15 | 24 | 12 | 11 | 12 |
| YCH931 | China | Han E | N1b1 | M178+, F3080+, F4063+, F3331+, F1419-, F4325- | 14 | 13 | 16 | 23 | 11 | 14 | 13 | 14 | 11 | 10 | 18 | 14 | 16 | 20 | 11 | 11 | 12 |
| YCH933 | China | Han E | N2b | F2569+ | 14 | 14 | 15 | 22 | 10 | 14 | 14 | 14 | 10 | 11 | 20 | 15 | 14 | 21 | 10 | 11 | 12 |
| YCH981 | China | Han E | N2a1* | F1833+, F846- | 14 | 14 | 17 | 23 | 10 | 12 | 13 | 14 | 10 | 11 | 19 | 16 | 15 | 23 | 12 | 11 | 12 |
| YCH1016 | China | Han E | N2a1* | F1833+, F846- | 14 | 13 | 17 | 23 | 10 | 14 | 13 | 14 | 11 | 11 | N/A | N/A | N/A | N/A | N/A | 12 | 12 |
| YCH1036 | China | Han E | N1* | F2130+, F3163-, M46- | 14 | 12 | 18 | 24 | 11 | 14 | 13 | 14 | 10 | 10 | N/A | N/A | N/A | N/A | N/A | 9 | 12 |
| YCH1038 | China | Han E | N1b1 | M178+, F3331?, F1419-, F4325- | 15 | 14 | 16 | 23 | 11 | 14 | 13 | 14 | 10 | 10 | N/A | N/A | N/A | N/A | N/A | 11 | 12 |
| YCH1039 | China | Han E | N1* | F2130+, F3163-, M46- | 14 | 13 | 15 | 24 | 11 | 15 | 13 | 14 | 10 | 11 | N/A | N/A | N/A | N/A | N/A | 11 | 12 |
| YCH1045 | China | Han N | N1b1a1* | F1419+, F4325+, F3271- | 13 | 14 | 16 | 24 | 10 | 16 | 14 | 14 | 11 | 12 | N/A | N/A | N/A | N/A | N/A | 11 | 13 |
| YCH1072 | China | Han E | N2a1* | F1833+, F846- | 14 | 14 | 16 | 23 | 10 | 14 | 14 | 14 | 10 | 12 | N/A | N/A | N/A | N/A | N/A | 11 | 12 |
| YCH1076 | China | Han E | N2b | F2569+ | 15 | 14 | 18 | 22 | 10 | 14 | 14 | 14 | 10 | 10 | N/A | N/A | N/A | N/A | N/A | 11 | 11 |
| YCH1077 | China | Han E | N1a1a1* | M128+, F866- | 15 | 14 | 16 | 23 | 10 | 15 | 13 | 14 | 10 | 11 | N/A | N/A | N/A | N/A | N/A | 12 | 13 |
| YCH1094 | China | Han E | N2a1* | F1833+, F846- | 14 | 14 | 17 | 23 | 10 | 14 | 14 | 14 | 10 | 11 | N/A | N/A | N/A | N/A | N/A | 11 | 13 |
| CAO-DC9 | China | Han S | N1a1a1* | M128+, F866- | 14 | 13 | 15 | 23 | 10 | 15 | 13 | 14 | 10 | 10 | 20 | 15 | 16 | 21 | 12 | 12 | 13 |
| CAO-DD0 | China | Han S | N1a1a1* | M128+, F866- | 14 | 12 | 15 | 23 | 10 | 15 | 13 | 14 | 10 | 11 | 20 | 15 | 16 | 21 | 12 | 12 | 13 |
| CAO-DD1 | China | Han S | N1a1a1* | M128+, F866- | 14 | 13 | 15 | 23 | 10 | 15 | 13 | 14 | 10 | 10 | 20 | 15 | 17 | 21 | 12 | 12 | 13 |
| CAO-PB1 | China | Han E | N2a1* | F1833+, F846- | 14 | 14 | 18 | 23 | 10 | 14 | 13 | 14 | 10 | 11 | 20 | 17 | 15 | 23 | 12 | 11 | 12 |
| CAO-NA1 | China | Han E | N1a1a1* | M128+, F866- | 14 | 13 | 16 | 23 | 10 | 15 | 13 | 14 | 10 | 11 | 19 | 16 | 16 | 21 | 12 | 12 | 12 |
| CAO-NA4 | China | Han E | N1a1a1* | M128+, F866- | 14 | 13 | 17 | 23 | 10 | 15 | 13 | 14 | 10 | 11 | 19 | 15 | 16 | 21 | 12 | 12 | 12 |
| CAO-AH0 | China | Han E | N1* | F2130+, F3163-, M46- | 14 | 13 | 16 | 23 | 10 | 15 | 13 | 14 | 10 | 11 | 19 | 16 | 17 | 21 | 12 | 13 | 14 |
| CAO-AT0 | China | Han E | N1b1* | M178+, F3331- | 14 | 14 | 16 | 23 | 10 | 13 | 13 | 14 | 10 | 10 | 18 | 14 | 17 | 21 | 11 | 11 | 12 |
| CAO-JE1 | China | Han E | N1a1a1* | M128+, F866- | 15 | 13 | 15 | 24 | 11 | 15 | 13 | 14 | 10 | 13 | 19 | 15 | 16 | 22 | 12 | 11 | 12 |
| HUI-AA7 | China | Hui | N1a1a1* | M128+, F866- | 14 | 14 | 16 | 23 | 10 | 14 | 14 | 15 | 10 | 10 | 18 | 14 | 18 | 21 | 11 | 11 | 12 |
| HLB-004 | China | MN, Ö, HB | N1a1* | F1154+, F2759- | 14 | 13 | 16 | 23 | 10 | 15 | 13 | 14 | 10 | 11 | 19 | 15 | 16 | 21 | 12 | 12 | 13 |
| HLB-022 | China | MN, BR, HB | N1* | F2130+, F3163-, M46- | 14 | 13 | 16 | 23 | 10 | 15 | 13 | 14 | 10 | 10 | 19 | 15 | 15 | 21 | 11 | 11 | 13 |
| HLB-026 | China | MN, BR, HB | N1b1a1a* | F3271+, F2288- | 14 | 14 | 16 | 23 | 10 | 14 | 14 | 14 | 10 | 10 | 19 | 14 | 20 | 22 | 12 | 11 | 13 |
| HLB-057 | China | MN, BR, HB | N1b1a1a* | F3271+, F2288- | 14 | 14 | 16 | 23 | 10 | 14 | 14 | 14 | 10 | 10 | 19 | 14 | 18 | 21 | 12 | 11 | 13 |
| HLB-060 | China | MN, BR, HB | N1b1a1a* | F3271+, F2288- | 14 | 14 | 16 | 23 | 10 | 14 | 14 | 14 | 10 | 10 | 19 | 14 | 18 | 22 | 12 | 11 | 13 |
| HLB-061 | China | MN, BR, HB | N1b1a1* | F1419+, F4325+, F3271- | 14 | 14 | 16 | 23 | 10 | 16 | 13 | 14 | 11 | 10 | 19 | 14 | 18 | 22 | 12 | 11 | 13 |
| HLB-066 | China | MN, BR, HB | N1b1a1a* | F3271+, F2288- | 14 | 14 | 16 | 23 | 10 | 14 | 13 | 14 | 10 | 10 | 19 | 14 | 18 | 22 | 12 | 11 | 13 |
| HLB-067 | China | MN, BR, HB | N1b1a1a* | F3271+, F2288- | 14 | 14 | 16 | 22 | 10 | 14 | 14 | 14 | 10 | 10 | 19 | 14 | 18 | 22 | 11 | 11 | 13 |
| HLB-068 | China | MN, BR, HB | N1b1a1a* | F3271+, F2288- | 14 | 14 | 16 | 22 | 10 | 14 | 14 | 14 | 10 | 10 | 19 | 14 | 18 | 22 | 11 | 11 | 13 |
| HLB-080 | China | MN, BR, HB | N1b1a1a* | F3271+, F2288- | 14 | 14 | 16 | 23 | 10 | 14 | 14 | 14 | 10 | 10 | 19 | 14 | 18 | 22 | 12 | 11 | 13 |
| HLB-082 | China | MN, BR, HB | N1b1a1* | F1419+, F4325+, F3271- | 14 | 14 | 16 | 23 | 11 | 14 | 14 | 14 | 10 | 10 | 19 | 15 | 17 | 22 | 12 | 11 | 11 |
| HLB-083 | China | MN, BR, HB | N1b1a1* | F1419+, F4325+, F3271- | 14 | 14 | 16 | 23 | 11 | 14 | 14 | 14 | 10 | 10 | 19 | 15 | 17 | 22 | 12 | 11 | 11 |
| HLB-084 | China | Dahur | N1* | F2130+, F3163-, M46- | 14 | 13 | 16 | 23 | 10 | 15 | 13 | 14 | 10 | 10 | 19 | 15 | 15 | 21 | 11 | 11 | 13 |
| HLB-085 | China | MN, BR, HB | N1b1a1a* | F3271+, F2288- | 14 | 14 | 16 | 23 | 10 | 14 | 14 | 14 | 10 | 10 | 19 | 14 | 18 | 22 | 12 | 11 | 13 |
| HLB-087 | China | MN, BR, HB | N1b1a1a1 | F2288+ | 14 | 14 | 16 | 23 | 10 | 14 | 14 | 14 | 10 | 10 | 19 | 14 | 17 | 22 | 12 | 11 | 13 |
| HLB-088 | China | MN, BR, HB | N1b1a1a* | F3271+, F2288- | 15 | 14 | 16 | 23 | 10 | 14 | 14 | 14 | 10 | 10 | 19 | 14 | 17 | 22 | 11 | 11 | 13 |
| HLB-089 | China | MN, BR, HB | N1b1a1a* | F3271+, F2288- | 15 | 14 | 16 | 23 | 10 | 14 | 14 | 14 | 10 | 10 | 19 | 14 | 17 | 22 | 11 | 11 | 13 |
| HLB-090 | China | MN, BR, HB | N1b1a1a* | F3271+, F2288- | 14 | 14 | 16 | 23 | 10 | 14 | 14 | 14 | 10 | 10 | 20 | 14 | 18 | 22 | 12 | 11 | 13 |
| HLB-091 | China | MN, BR, HB | N1b1a1a* | F3271+, F2288- | 14 | 14 | 16 | 23 | 10 | 14 | 14 | 14 | 10 | 10 | 19 | 14 | 18 | 21 | 12 | 11 | 13 |
| HLB-092 | China | MN, BR, HB | N1b1a1a* | F3271+, F2288- | 14 | 14 | 16 | 23 | 10 | 14 | 14 | 14 | 10 | 10 | 19 | 14 | 18 | 22 | 12 | 13 | 13 |
| HLB-093 | China | MN, BR, HB | N1b1a1a* | F3271+, F2288- | 14 | 14 | 16 | 23 | 10 | 14 | 14 | 14 | 10 | 10 | 19 | 14 | 18 | 22 | 12 | 11 | 13 |
| HLB-098 | China | MN, BR, HB | N1b1a1a* | F3271+, F2288- | 14 | 14 | 16 | 23 | 10 | 14 | 14 | 14 | 10 | 10 | 19 | 14 | 18 | 22 | 12 | 11 | 13 |
| HLB-101 | China | MN, BR, HB | N1b1a1a | F3271+ | 14 | 13 | 17 | 23 | 10 | 14 | 14 | 14 | 10 | 10 | 19 | 15 | ? | 22 | 12 | 11 | 13 |
| HLB-102 | China | MN, BR, HB | N1b1a1* | F1419+, F4325+, F3271- | 14 | 14 | 16 | 23 | 10 | 16 | 13 | 14 | 11 | 10 | 19 | 14 | 18 | 22 | 12 | 11 | 13 |
| HLB-105 | China | MN, BR, HB | N1b1a1a* | F3271+, F2288- | 14 | 14 | 16 | 23 | 10 | 14 | 14 | 14 | 10 | 10 | 19 | 14 | 18 | 21 | 12 | 11 | 13 |
| HLB-106 | China | MN, BR, HB | N1b1a1a* | F3271+, F2288- | 14 | 14 | 17 | 23 | 10 | 14 | 15 | 14 | 10 | 10 | 19 | 14 | 18 | 22 | 12 | 11 | 13 |
| HLB-107 | China | MN, BR, HB | N1b1a1a* | F3271+, F2288- | 14 | 14 | 16 | 23 | 10 | 14 | 14 | 14 | 10 | 10 | 19 | 14 | 18 | 22 | 12 | 11 | 13 |
| HLB-108 | China | MN, BR, HB | N1b1a | F3331+, F3271- | 15 | 13 | 16 | 24 | 10 | 16 | 14 | 14 | 10 | 12 | 18 | 15 | 18 | 23 | 10 | 14 | 19 |
| HLB-111 | China | MN, BR, HB | N1b1a1a* | F3271+, F2288- | 14 | 14 | 16 | 23 | 10 | 14 | 14 | 14 | 10 | 10 | 19 | 14 | 19 | 22 | 12 | 11 | 13 |
| HLB-113 | China | MN, BR, HB | N1b1a1a* | F3271+, F2288- | 14 | 14 | 16 | 23 | 10 | 14 | 14 | 14 | 10 | 10 | 19 | 14 | 18 | 22 | 12 | 11 | 13 |
| HLB-114 | China | MN, BR, HB | N1b1a1* | F1419+, F4325+, F3271- | 14 | 14 | 16 | 23 | 10 | 16 | 13 | 14 | 11 | 10 | 19 | 14 | 18 | 22 | 12 | 11 | 13 |
| HLB-118 | China | MN, BR, HB | N1b1a1a* | F3271+, F2288- | 14 | 14 | 16 | 23 | 10 | 14 | 14 | 14 | 10 | 10 | 19 | 14 | 19 | 21 | 12 | 11 | 13 |
| HLB-121 | China | Evenk | N1* | F2130+, F3163-, M46- | 14 | ? | 16 | 23 | 10 | 14 | 13 | 14 | 10 | 10 | 19 | 15 | 16 | 24 | 12 | 12 | 13 |
| HLB-131 | China | MN, Ö, HB | N1b1 | M178+ | 14 | 14 | 16 | 23 | 10 | 14 | 12 | 14 | 9 | 11 | 18 | 14 | 19 | 21 | 11 | 11 | 12 |
| HLB-148 | China | MN, BR, HB | N1b1a1a* | F3271+, F2288- | 14 | 14 | 16 | 23 | 10 | 14 | 13 | 14 | 10 | 10 | 19 | 14 | 18 | 22 | 12 | 11 | 13 |
| HLB-155 | China | MN, BR, HB | N1b1a1a* | F3271+, F2288- | 14 | 14 | 16 | 23 | 10 | 14 | 14 | 14 | 10 | 10 | 19 | 14 | 17 | 22 | 12 | 11 | 13 |
| HLB-157 | China | MN, BR, HB | N1b1a1a* | F3271+, F2288- | 14 | 13 | 17 | 23 | 10 | 14 | 14 | 14 | 10 | 10 | 19 | 14 | 18 | 22 | 12 | 11 | 13 |
| HLB-161 | China | MN, BG, HB | N1a1a* | F2759+, M128- | 14 | 14 | 16 | 24 | 10 | 14 | 13 | 14 | 10 | 10 | 18 | 15 | 16 | 25 | 12 | 12 | 13 |
| HLB-163 | China | MN | N1* | F2130+, F3163-, M46- | 14 | 13 | 15 | 24 | 11 | 15 | 13 | 14 | 10 | 11 | 19 | 16 | 16 | 21 | 13 | 11 | 12 |
| HLB-165 | China | MN, BG, HB | N1b1a1a1 | F2288+ | 14 | 14 | 16 | 23 | 10 | 14 | 14 | 14 | 10 | 10 | 19 | 14 | 17 | 22 | 13 | 11 | 13 |
| HLB-176 | China | MN, BG, HB | N1b1a1a* | F3271+, F2228- | 14 | 14 | 16 | 23 | 10 | 14 | 14 | 14 | 10 | 10 | 19 | 14 | 18 | 22 | 12 | 11 | 13 |
| HLB-185 | China | MN, BG, HB | N1b1a1a1 | F2288+ | 14 | 13 | 17 | 22 | 9 | 14 | 14 | 14 | 10 | 11 | 19 | 16 | 17 | 22 | 10 | 11 | 14 |